\documentclass[epj]{svjour}
\usepackage{amsmath}
\usepackage{amssymb}
\usepackage{natbib}
\setcitestyle{numbers}
\setcitestyle{square}
\usepackage{hyperref}
\usepackage{graphicx}
\usepackage{color}
\usepackage{dcolumn}
\usepackage{bm}
\usepackage[T1]{fontenc}
\usepackage{url}

\definecolor{OliveGreen}{rgb}{0,0.6,0}

\definecolor{orange}{rgb}{1,0.5,0}

\begin{document}
\title{On the Deconfinement Phase Transition in Neutron-Star Mergers}
\author{Elias R. Most \inst{1,2} \and L. Jens Papenfort \inst{1} \and
  Veronica Dexheimer \inst{3} \and Matthias Hanauske \inst{1,4} \and
  Horst Stoecker \inst{1,4,5} \and Luciano Rezzolla \inst{1,6}
}                     
\offprints{}          
\institute{{Institut f{\"u}r Theoretische Physik, Max-von-Laue-Stra{\ss}e
    1, 60438 Frankfurt, Germany} \and
  Center for Computational Astrophysics, Flatiron Institute, 
  162 Fifth Avenue, New York, NY 10010, USA
  \and Department of Physics, Kent State
  University, Kent, OH 44243 USA \and Frankfurt Institute for Advanced
  Studies, Ruth-Moufang-Stra{\ss}e 1, 60438 Frankfurt, Germany \and GSI
  Helmholtzzentrum f{\"u}r Schwerionenforschung GmbH, 64291 Darmstadt,
  Germany \and School of Mathematics, Trinity College, Dublin 2, Ireland}
\date{Received: date / Revised version: date}
%
\abstract{We study in detail the nuclear aspects of a neutron-star merger
  in which deconfinement to quark matter takes place. For this purpose,
  we make use of the Chiral Mean Field (CMF) model, an effective
  relativistic model that includes self-consistent chiral symmetry
  restoration and deconfinement to quark matter and, for this reason,
  predicts the existence of different degrees of freedom depending on the
  local density/chemical potential and temperature. We then use the
  out-of-chemical-equilibrium finite-temperature CMF equation of state in
  full general-relativistic simulations to analyze which regions of
  different QCD phase diagrams are probed and which conditions, such as
  strangeness and entropy, are generated when a strong first-order phase
  transition appears. We also investigate the amount of electrons
  present in different stages of the merger and discuss how far from
  chemical equilibrium they can be and, finally, draw some comparisons
  with matter created in supernova explosions and heavy-ion collisions.
\PACS{
      {PACS-key}{discribing text of that key}   \and
      {PACS-key}{discribing text of that key}
     } 
} 
\maketitle
\section{Introduction}

The interior of neutron stars covers an incredible range of densities
going from about $10^4$ g/cm$^3$ in the crust to about $10^{15}$ g/cm$^3$
in the core, corresponding to several times the nuclear saturation
density \cite{Rezzolla2018}. During a neutron-star merger this value can
increase to several times $10^{15}$ g/cm$^3$ in the center, corresponding
to more than $10$ times the nuclear saturation density (see
Refs.~\cite{Baiotti2016,Paschalidis2016} for some recent reviews). Such
extreme densities combined with temperatures of several tens of MeV are
particularly relevant if the equation of state (EOS) allows for a
deconfinement to quark matter takes place \cite{Oechslin:2004, Most2018b,
  Bauswein:2018bma}. Clearly, the investigation of these scenarios
requires, from one hand, the use of accurate numerical-relativity
calculations and, from the other hand, a microscopic description that
allows for the existence exotic degrees of freedom, such as hyperons and
quarks (see Ref.~\cite{Baiotti:2019sew} for a review of the relation
between gravitational waves and the microscopic description of neutron
stars).

It has been shown in Refs.~\cite{Sekiguchi2011b,Radice2017a}
that hyperons can modify the
frequency and amplitude of gravitational waves emitted by neutron-star
mergers. These changes are expected to be visible even before the merger
takes place, as hyperons are usually triggered at intermediate densities,
specially when temperature effects are pronounced. Deconfinement to quark
matter, on the other hand, was found to modify the frequency and
amplitude of gravitational waves emitted only at \cite{Bauswein:2018bma}
or after the merger \cite{Most2018b}. The possibility of the merger of
pure quark stars have also investigated in the past \cite{Bauswein2009}. 
More recently, attention has been paid to the merger of twin stars
in terms of equilibrium models \cite{Montana2018, Zhang2019d} or through
simulations in full general relativity of the merger of a hadronic and a quark star \cite{DePietri2019} and quark stars with hadronic crusts \cite{Gieg2019}.
In a holographic approach, Ref~\cite{Ecker2019} has found not to be possible to
reach the phase transition to the quark phase before collapsing to a black
hole.  The complex dynamics found in these works, as well as the impact on
the electromagnetic counterpart to be expected from this process
\cite{Bucciantini2019}, clearly calls for more extended and detailed work.

It has been shown that a deconfinement phase transition can produce shock waves in stars \cite{Mishustin2015}. In a previous work \cite{Most2018b}, we reported that a strong first-order phase transition
to quark matter can lead to a post-merger gravitation-al-wave signal that
is different from the one expected from the inspiral, which can only
probe the hadronic part of the EOS. In particular, within the scenario
investigated in Ref.~\cite{Most2018b}, small amounts of quarks in hot
regions of the hypermassive neutron star (HMNS) lead to a dephasing of
the signal, while the appearance of a strong first-order phase transition
induces an early collapse of the remnant to a black hole, producing a
ringdown signal which is different from the collapse of a purely hadronic
remnant. Here, we provide a number of additional pieces of information
and expand on the analysis carried out in Ref.~\cite{Most2018b}. In
particular, we here focus on the nuclear aspects of a merger event in
which a deconfinement phase transition takes place in order to understand
how the outcome compares to matter generated in core-collapse supernova
explosions or in heavy-ion collisions. For this, we analyze the light and
strange-quark content at the time when a hot quark-phase is formed in the
HMNS. We also show what thermodynamical conditions, such as temperature
and entropy, and compositions (charge, lepton, and strangeness fractions)
are generated and which baryon, charge, and electron chemical potentials
they correspond to. This can serve as a guide for nuclear physicists who
want to study the effects of neutron-star merger conditions in their EOSs
that contain exotic degrees of freedom.

The plan of the paper is as follows: first, we discuss the microscopic
EoS and the hydrodynamical code used for the merger simulation. Then, we
present the outcome of our simulations and discuss our results. Finally,
we compare our findings with other hot and dense environments and present
our conclusions.

\section{Methods}

\subsection{Equation of state}

Due to the extreme conditions expected to be found in neutron-star
mergers, it is compelling to construct the EOS applying a formalism that
includes the basic features predicted by QCD, namely chiral-symmetry
restoration and quark deconfinement. In the absence of a fundamental
theory that can be applied in the whole energy regime and conditions
necessary for our study, we make use of an effective model, the Chiral
Mean Field (CMF) model, which is based on a nonlinear realization of the
SU(3) chiral sigma formalism \cite{Papazoglou:1998}. It is a relativistic
model constructed from symmetry relations, which allows it to be chirally
invariant in the expected regime. The baryon and quark masses are
generated by interactions with the medium and, therefore, decrease with
temperature and or chemical potential/density. The Lagrangian density of
the CMF model in the mean field approximation reads \cite{Dexheimer:2008,
  Dexheimer:2009}
\begin{equation}
  \label{eq:lagrangian}
  L = L_{\rm kin} + L_{\rm int} + L_{\rm self}+ L_{\rm sb}
  - U,
\end{equation}
where, besides the kinetic energy term for hadrons, quarks and electrons
($L_{\rm kin}$), the terms remaining correspond to the interaction
between the octet of baryons wit spin $1/2$, the $3$ lighter quarks, and
mesons ($L_{\rm int}$), self interactions of scalar and vector mesons
($L_{\rm self}$), an explicit chiral symmetry breaking term necessary to
produce vacuum masses for the pseudo-scalar mesons ($L_{\rm sb}$), and
the effective potential $U$ for the scalar field $\Phi$. This scalar
field was named in an analogy to the Polyakov loop in the (Polyakov)
Nambu and Jona-Lasinio (PNJL) approach \cite{Ratti:2006ka,
  Roessner:2006xn} and its potential in our approach depends on the
temperature $T$ and the baryon chemical potential $\mu_B$
\begin{eqnarray} U &=& (a_o T^4 + a_1 \mu_B^4 + a_2 T^2 \mu_B^2) \Phi^2
  \nonumber \\ &+& a_3 T_o^4 \ \ln{(1 - 6 \Phi^2 + 8 \Phi^3 -3 \Phi^4)} .
\end{eqnarray}

The mesons included are the vector-isoscalars $\omega$ and $\phi$ (strange
quark-antiquark state), the vector-isovector $\rho$, the scalar-isoscalars
$\sigma$ and $\zeta$ (also strange quark-antiquark state), and the
scalar-isovector $\delta$. They are treated as classical fields within the
mean-field approximation. Finite-temperature calculations include the heat
bath of hadronic and quark quasiparticles within the grand canonical
ensemble. The grand potential density of the system is defined as
\begin{eqnarray}
  \label{eq:grand_potential}
  \frac{\Omega}{V} &=& -L_{\rm int} - L_{\rm self} - L_{\rm
  sb} - L_{\rm vac} + U \nonumber\\ &+& T \sum_i \frac{\gamma_i}{(2 \pi)^3}
  \int_{0}^{\infty} \, d^3k \, \ln(1 + e^{-\frac{1}{T}(E_i^*(k) \mp
    \mu_i^*)} ),
\end{eqnarray}
where $L_{\rm vac}$ is the vacuum energy, $\gamma_i$ is the fermionic
degeneracy (which for the quarks also includes color degeneracy),
$E_{i}^* (k) = \sqrt{k_i^2 + {M^*_i}^2}$, is the single particle
effective energy and
\begin{eqnarray} \mu_i^* = \mu_i - g_{i\omega} \omega - g_{\phi} \phi -
  g_{i\rho} \tau_3 \rho , \end{eqnarray}
is the effective chemical potential of each species. The $-$ and $+$ signs in
the grand potential density \eqref{eq:grand_potential} refer to
particles and antiparticles, respectively.  The chemical potential for
each species $\mu_i$ is determined by the conditions imposed to the
system, conserved baryon number and electric charge,
\begin{eqnarray} \mu_i = Q_{B,i} \, \mu_B + Q_i \,\mu_Q ,
  \label{mu_i}
\end{eqnarray}
where $\mu_B$ and  $\mu_Q$ represent the chemical potentials corresponding
to the conserved quantities and the values $Q_{B,i}$ and $Q_i$ are the
baryon charge ($1$ for baryons and $1/3$  for quarks) and electric charge
of a particular species $i$. 

The coupling constants of the hadronic sector of the model were fitted to
reproduce vacuum masses of baryons and mesons, nuclear saturation
properties (density $\rho_0=0.15$ fm$^{-3}$, binding energy per nucleon
$B/A=-16$ MeV, and compressibility $K=300$ MeV), the asymmetry energy
($E_{sym}=30$ MeV with slope $L=88$ MeV), and reasonable values for the
hyperon potentials ($U_\Lambda=-28.00$ MeV, $U_\Sigma=5$ MeV, and
$U_\Xi=-18$ MeV). The reproduced critical point for the nuclear
liquid-gas phase transition lies at $T_c=16.4$ MeV, $\mu_{B,c}=910$ MeV,
while the vacuum expectation values of the scalar mesons are constrained
by reproducing the pion and kaon decay constants.

Due to their interactions with the mean field of mesons and the field
$\Phi$, the effective masses of baryons and quarks take the following form in our approach
\begin{eqnarray} M_{B}^* &=& g_{B \sigma} \sigma + g_{B \delta} \tau_3
  \delta + g_{B \zeta} \zeta + M_{0_B} + g_{B \Phi} \Phi^2, 
  \\
M_{q}^* &=& g_{q \sigma} \sigma + g_{q \delta} \tau_3 \delta + g_{q \zeta}
\zeta + M_{0_q} + g_{q \Phi}(1 - \Phi), \end{eqnarray}
where $M_0$ are small bare-mass terms. Notice that for low values of
$\Phi$, $M_{B}^*$ is small while $M_{q}^*$ is very large. This
essentially indicates that, for low $\Phi$, the presence of baryons is
promoted while quarks are suppressed, and vice versa. In this sense,
$\Phi$ acts as an order parameter for deconfinement. The potential $U$,
together with the quark couplings, has been fit to reproduce several
features expected from the QCD phase diagram, including lattice data for
pure gauge and with quarks (procedure explained in detail in
Ref.~\cite{Dexheimer:2009}). In the latter case, we reproduce a crossover
at vanishing and small chemical potential, after which a first-order
coexistence line starts, continuing all the way to the zero temperature
axis. The values of all coupling constants can be found in
Ref.~\cite{Roark2018}.

It should be mentioned that the CMF model allows for the existence of
soluted quarks in the hadronic phase and soluted hadrons in the quark
phase at finite temperature. This is different from a Gibbs construction
and the appearance of mixture of phases, which eliminates discontinuities
in the first derivatives of the grand potential (see
Refs.~\cite{Glendenning:1992vb,Lukacs:1986hu,
  Heinz:1987sj,Poberezhnyuk:2018mwt} and Ref.~\cite{Hempel:2013} with
references therein for details). Regardless, quarks always give the
dominant contribution in the quark phase, and hadrons in the hadronic
phase. We assume that this inter-penetration of quarks and hadrons is
indeed physical, and is required to achieve the crossover transition,
known to take place at low $\mu_B$ values \cite{Aoki:2006}.

For cold chemically equilibrated matter, the formalism leads to a neutron
star with maximum mass of $2.07\,M_\odot$ and a radius of $12$ km when
quarks are suppressed. Otherwise, the model presents a very strong
deconfinement phase transition that destabilizes stars (as no
quark-vector interactions are included in agreement with lattice QCD
\cite{Steinheimer2014}), unless a mixtures of phases is allowed. In the
latter case, we reproduce a stable maximum-mass star with $1.93\,M_\odot$
and a radius of $13$ km \cite{Roark2018}, more than $2$ km of which
contain quarks. For the canonical star with mass $1.4\,M_\odot$, a
corresponding radius of $14$ km is found. In addition, when we consider
nonlinear isovector singlet to isovector triplet coupling of the vector
mesons for the baryons, the radius of the $1.4\,M_\odot$ star reduces to
less than $13$ km \cite{Dexheimer:2018dhb}. These values of the maximum
mass and radii are compatible with the expectations matured after the
first detection of gravitational waves from a binary neutron-star merger
(GW170817) \cite{Abbott2017_etal, Annala2017, Bauswein2017b,
  Margalit2017, Radice2017b, Rezzolla2017, Ruiz2017, Shibata2017c,
  Most2018}. For pro-neutron stars, we reproduce a stable maximum-mass
star with $2.03\,M_\odot$ and a radius of $18$ km, more than $7$ km of
which contain quarks \cite{Roark2018}.

In order to use our microscopic formalism in neutron-star
merger simulations, we build 3-dimensional tables in which we vary the
baryon number density, charge fraction, and temperature. The baryon number
density is defined as 
\begin{eqnarray} n_B=\left. -\frac{\partial \Omega/V}{\partial
  \mu_B}\right|_{T, V,\mu_Q}=\sum_i  n_i -\frac{\partial U}{\partial
  \mu_B}, \end{eqnarray}
where $n_i$ is the number density of particle species $i$. The extra
contribution of the gluons to the baryon density represents color bound
states and mimics extra possible states, as for example the contribution
of higher resonances.

The charge fraction is calculated as the amount of electric charge per
baryon (and quark) and it is only summed over baryons and quarks
\begin{eqnarray} Y_Q = \frac{Q}{B} = \frac{\sum_i Q_i \ n_i}{\sum_i Q_{B,i}
\ n_i}. \end{eqnarray}
The electrons, which are not considered to be in chemical equilibrium with
the rest of the system, are then added in order to fulfill electric charge
neutrality 
\begin{eqnarray} n_e= -{\sum_i  Q_i \ n_i} = Y_Q \sum_i Q_{B,i} \ n_i.
\end{eqnarray}

In Ref.~\cite{Dexheimer:2017nse}, it was described in detail the
construction of the three-dimensional table, which is already available
online on the CompOSE repository \cite{Typel2015,compose} for hadronic
matter. In the future, equivalent tables but that also contain quark
degrees of freedom will be uploaded.  Figures (3) and (4) of
Ref.~\cite{Dexheimer:2017nse} show the effect of the increase of charge
fraction in the CMF model. At zero temperature, going from $Y_Q=0$ to
$Y_Q=0.5$ essentially eliminates hyperons and pushes strange quarks to
densities that are too high to be important for cold neutron stars.

\subsection{Numerical infrastructure}


For completeness, we quickly summarize the numerical methods used for the
simulations reported here and first discussed in
Ref.~\cite{Most2018b}. More specifically, we solve a coupled system of
the Einstein and general-relativistic magnetohydrodynamics (GRMHD)
equations using the code \texttt{Frankfurt/IllinoisGRMHD} (\texttt{FIL}),
which is a high-order extension of the publicly available
\texttt{IllinoisGRMHD} code \citep{Etienne2015} part of the
\texttt{Einstein Toolkit} \citep{loeffler_2011_et}. In the following, we
give an overview of the numerical details and implementation of the
formalism.

To solve the Einstein equations, \texttt{FIL} provides its own spacetime
evolution module, which implements the Z4c
\citep{Hilditch2012,Bernuzzi:2009ex} and CCZ4 \citep{Alic:2011a,Alic2013}
formulations using forth-order accurate finite differencing
\citep{Zlochower2005:fourth-order} with different choices for the
conformal factor. In this work, we choose $\psi^{-2}$ and adopt the Z4c
formulation with a damping coefficient $\kappa=0.02$
\citep{Weyhausen:2011cg, Hilditch2012}. The space-time gauges are evolved
using the standard 1+log slicing and shifting-shift Gamma driver
conditions \citep{Alcubierre:2008, Rezzolla_book:2013}, where a uniform
damping parameter of $\eta = 2/M$ is adopted.

The GRMHD equations are solved using the ECHO scheme \citep{DelZanna2007},
making our code overall formally fourth-order accurate. The fluxes are
computed from the reconstructed primitive variables using a HLLE Riemann
solver \citep{Harten83}. The reconstruction step $\mathcal{R}$ is
performed using the WENO-Z method \citep{Borges2008}, with the optimal
weights and stencils for a conservative finite difference scheme taken
from Ref.~\citep{DelZanna2007}.

\begin{figure*}[t]
  \includegraphics[width=0.95\columnwidth]{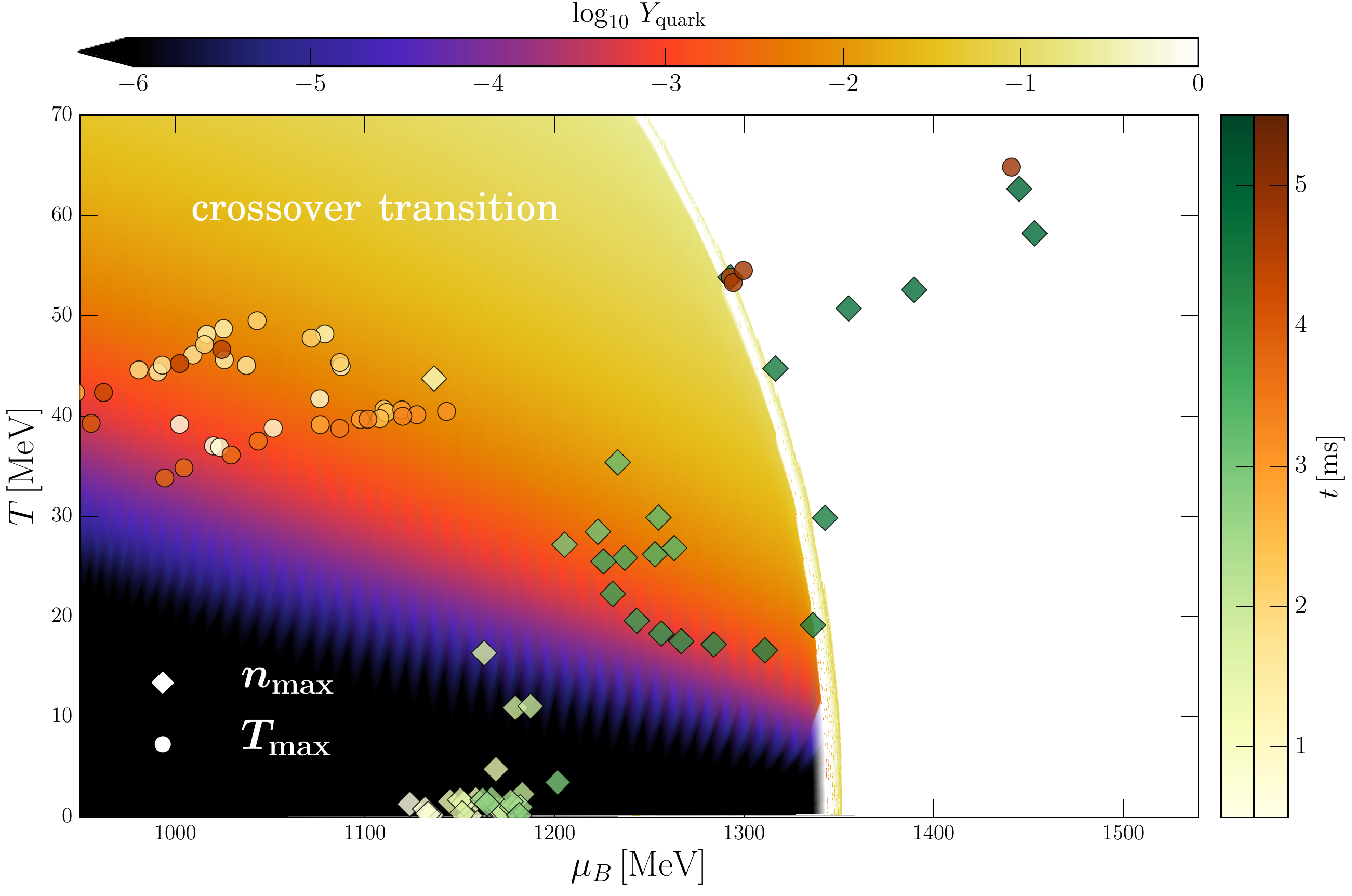}
  \hspace{10mm}
  \includegraphics[width=0.95\columnwidth]{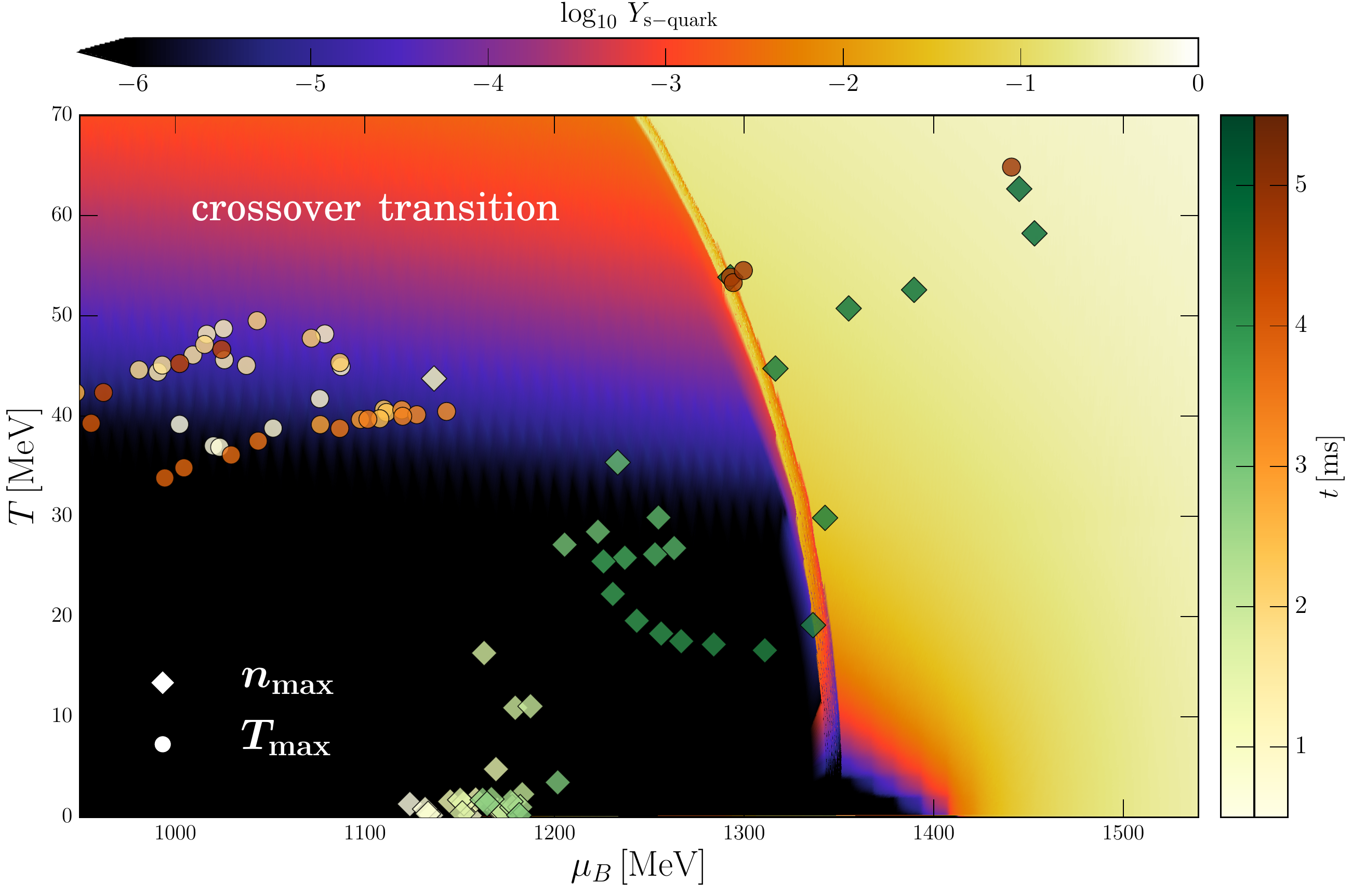}
  \caption{\label{T-mub} \textit{Left panel:} Evolution of the densest
    and hottest parts of the hypermassive neutron star in the QCD phase diagram.
    The background color refers to the total quark fraction predicted by
    the CMF model for charge fraction $Y_Q=0.05$. The different symbols
    describe the evolution of the largest baryon density and largest
    temperature points encountered during the simulation. \textit{Right
      panel:} Same as in the left panel, but showing in the background
    the strange-quark fraction predicted by the CMF model for charge
    fraction $Y_Q=0.05$.}
\end{figure*}

Our initial data is represented by an irrotational equal-mass neutron
star binary with a total mass of $2.9\,M_\odot$ constructed using the
\texttt{LORENE} code having an initial proper separation of $45\,\rm
km$. The simulation domain is modeled by a series of seven nested boxes
extending up to $\simeq 1500\,{\rm km}$ for which the finest-grid box has
a resolution of $250\, \rm m$.

\section{Results}

We start by analyzing the evolution of the densest and hottest parts of
the HMNS in the left panel of Fig.~\ref{T-mub}, which reports
the regions of the standard QCD phase diagram (temperature vs$.$ baryon
chemical potential) that are probed in our neutron-star merger simulation. In
particular, we show with a color-code in the background of the figure the
quark fraction $Y_{\rm quark}$ (i.e., the number of quarks normalized by
the total number of baryons and quarks $B$) predicted by CMF model when
the charge fraction for the baryons and quarks is fixed to
$Y_Q=0.05$. This is a good approximation for the charge fraction present
at intermediate densities when the merger event starts (see discussion
below for Fig.~\ref{Yq-mub}). As expected from our formalism, the phase
transition is quite sharp at zero and low temperatures, reproducing pure
hadronic matter to the left (black region) and pure quark matter to the right (white region) of the
coexistence line. On the other hand, the phase transition becomes
smoother for larger temperatures to the point that, if we had extended
the figure to larger temperatures, the first-order coexistence line would
had disappeared before reaching the zero baryon chemical potential axis,
at a critical temperature of $T_c=169$ MeV.

The finite width for the coexistence line in the left panel of
Fig.~\ref{T-mub} is related to the use of the baryon chemical potential
for the horizontal axis, which is not the Gibbs free energy per baryon of
the system in the case of a fixed charge fraction. The independent
chemical potential and Gibbs free energy per baryon of the system in this
case is $\tilde{\mu} = \mu_B + Y_Q \mu_Q$ \cite{Hempel:2013,Typel2015}.

The different symbols in Fig.~\ref{T-mub} describe the evolution of the
largest baryon density and largest temperature points encountered during
the simulation. The largest density represented by diamonds corresponds,
first, to the reminiscent of the original neutron stars and, later, to
the center of the HMNS formed by the merger. As time evolves, these
points correspond to larger baryon chemical potentials and, on average,
larger temperatures, except for the earlier stage when the densest points
switch between the two reminiscent stars.  After $\simeq 4.5\, \rm ms$ ,
the coexistence line is crossed and a large amount of deconfined quark
matter appears in the center of the newly formed HMNS. The hottest region
represented by circles on the figure corresponds to different regions of
the merger and, only after deconfinement to quark matter takes place, it
coincides with the center of the HMNS (brown circles). Before that, the
hottest region appears off-center in the shape of a ring
\cite{Kastaun2016, Hanauske2016}, mostly as a manifestation of the
conservation of the Bernoulli constant \cite{Hanauske2016}.

The right panel of Fig.~\ref{T-mub} also shows which regions of the
standard QCD phase diagram are probed in a neutron-star merger but, now,
we show in the background the amount of strange quarks $Y_{\rm s-quark}$
(number of s-quarks normalized by $B$) predicted by the CMF model for
$Y_Q=0.05$. Note that strange quarks do not appear immediately after the
deconfinement phase transition at low temperatures (black region), a consequence of
their large bare mass.  Even at intermediate and larger temperatures, the
amount of strange quarks is roughly an order of magnitude less than that
of up and down quarks combined. Nevertheless, in the final stages of the
merger, a combination of large densities and temperatures can produce
$Y_{\rm s-quark} > 10\%$ in the center of the HMNS.

\begin{figure}[t!] 
  \includegraphics[width=8.7cm, trim=0 0 0 0,clip]{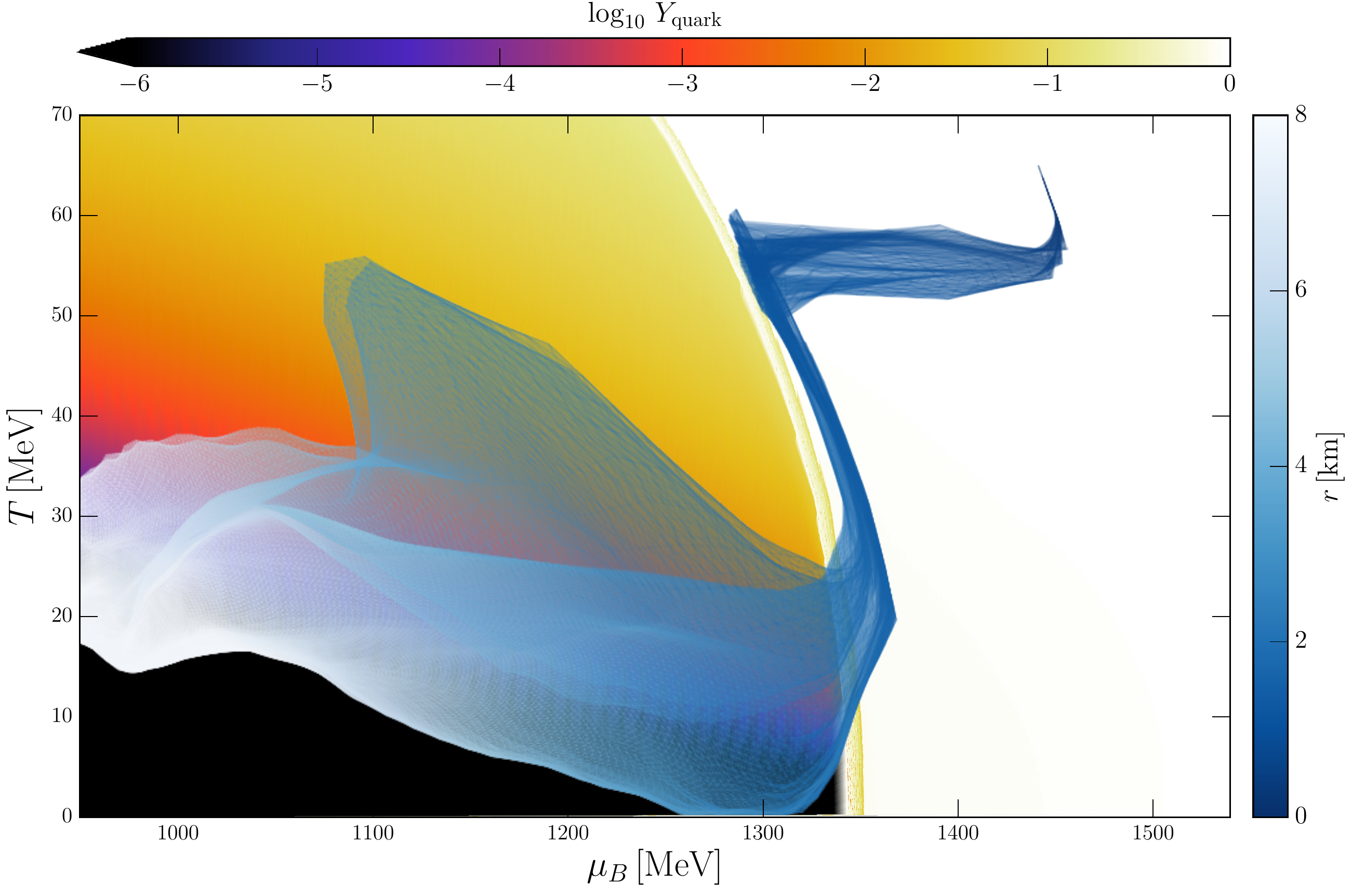}
  \caption{\label{T-mubv1} Portion of the QCD phase diagram covered by
    the simulation immediately after deconfinement to quark
    matter has taken place in its center. The background
    color is the same as in the left panel of Fig. \ref{T-mub}. The blue
    scale shows the distance to the center of the hypermassive neutron
    star.}
\end{figure}

To better highlight the various regions of the phase diagram probed by
the HMNS, Fig.~\ref{T-mubv1} shows the temperature and baryon chemical
potential values covered at a fiducial time when the quark phase has
already formed after the merger, about $5\, \rm ms$ for the case a total
mass of $2.9\, M_\odot$.  The blue color code shows the distance to the
center of the HMNS up to $8$ km of radius. We can see that the hadronic
part of the HMNS covers a large area of the phase diagram extending up to
$T>55\, \rm MeV$ \cite{hanauske2019a,hanauske2019b}. In the region where
the deconfinement takes place, there is a large temperature increase
related to the gravitational collapse due to the softening of the EoS
across the first-order phase transition. The softening is related to the
extent of the energy or baryon number density jump across the phase
transition, the latter having already being shown in the horizontal axis
of Fig.~3 of Ref.~\cite{Most2018b}. Note that our deconfinement phase
transition is not an adiabatic process. During this stage, even if the
temperature was kept constant, the entropy $S_B$ would increase by a
factor $\sim3$, related to the appearance of color degrees of freedom and
different interactions in the quark phase. In the deconfined phase, the
temperature reaches even larger values $T\simeq 60\, \rm MeV$. Beyond the
rightmost point of the phase diagram, an apparent horizon starts to form
and, as the simulation proceeds, the HMNS collapses to black hole in a
few ms.

\begin{figure*}
  \includegraphics[width=0.95\columnwidth]{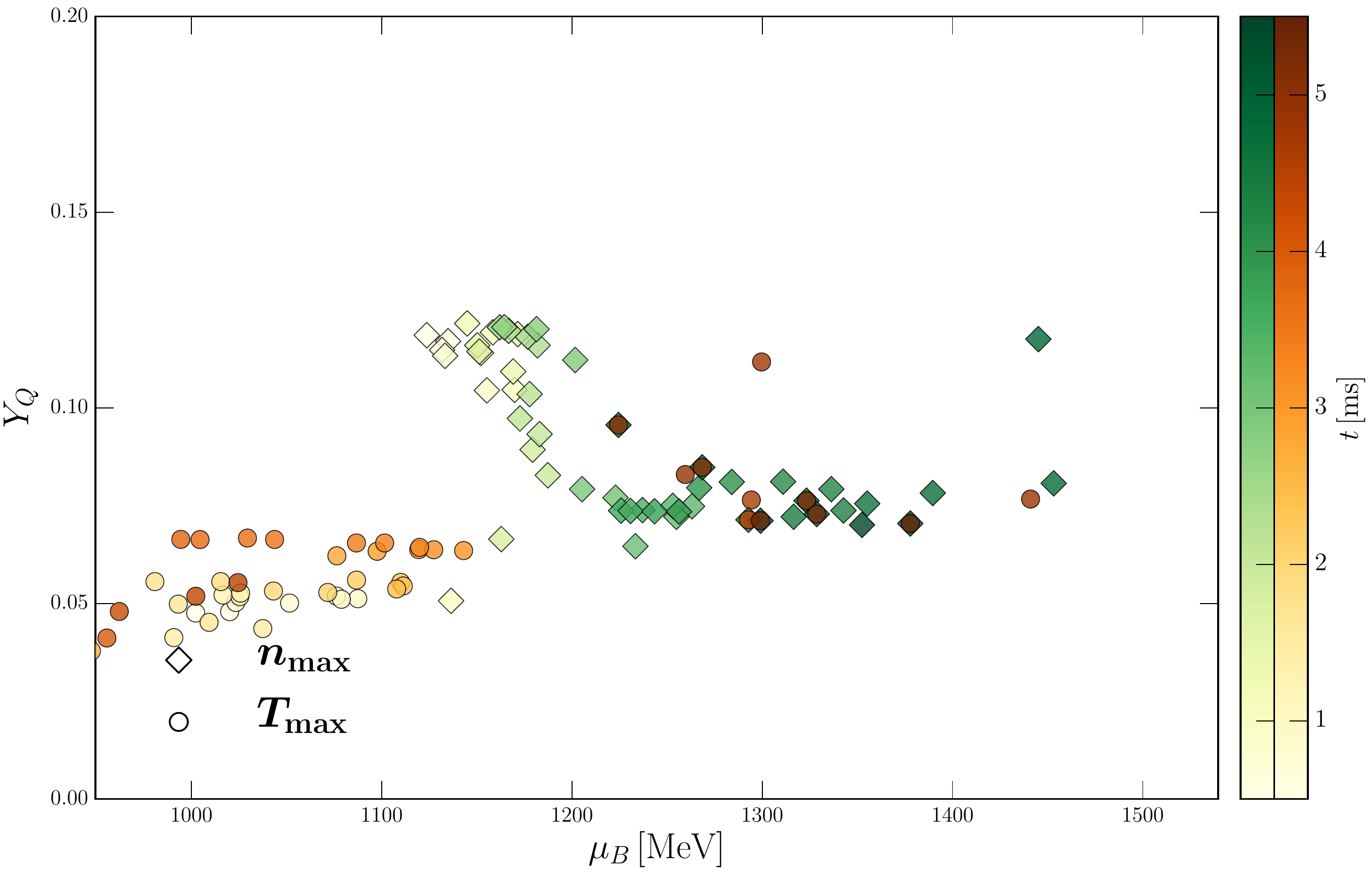}
  \hspace{10mm}
  \includegraphics[width=0.95\columnwidth]{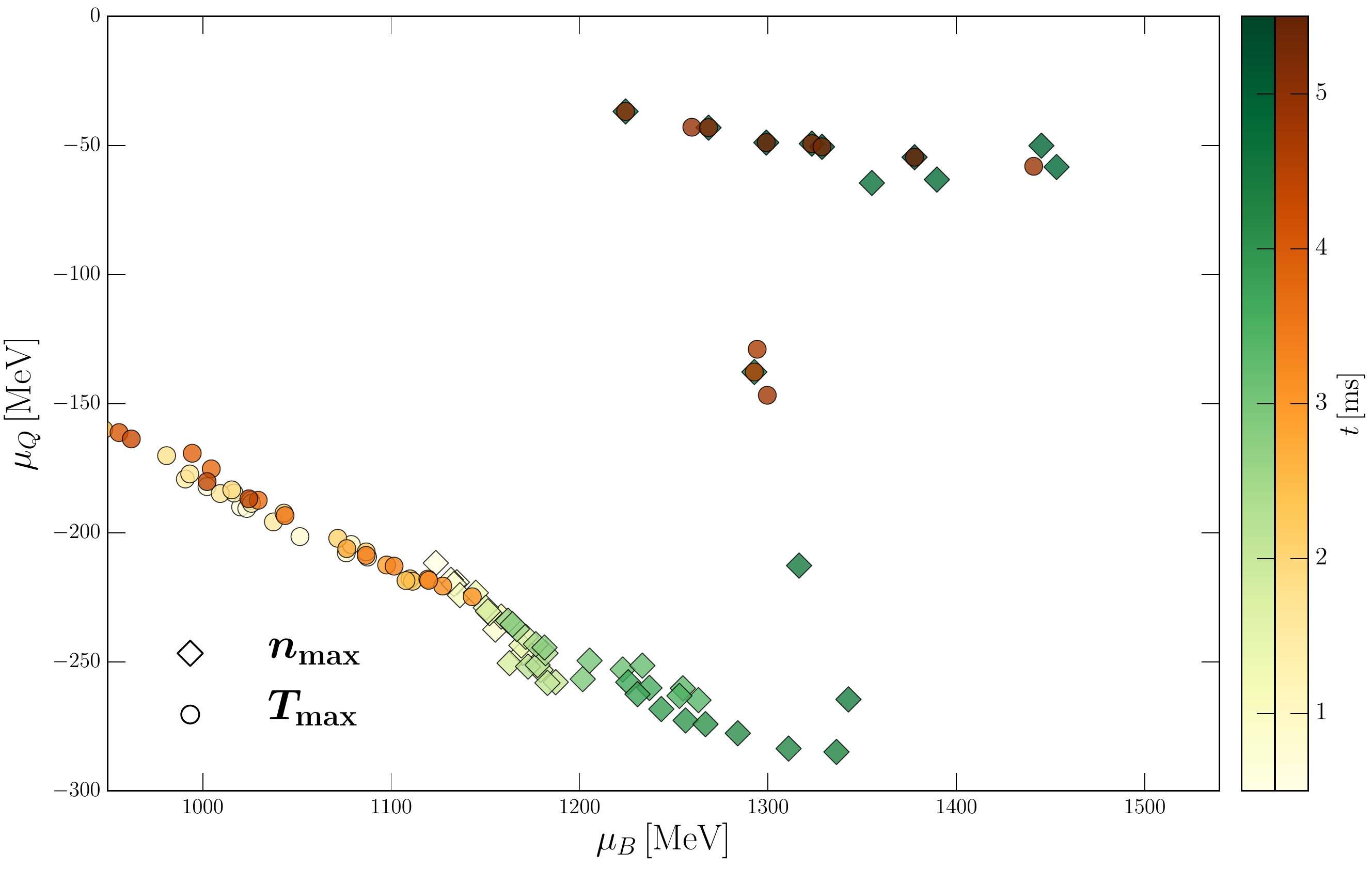}
  \caption{\label{Yq-mub} \textit{Left panel:} The charge fraction vs$.$
    baryon chemical potential phase diagram. The symbols once more follow
    the densest and hottest points of the hypermassive neutron star. \textit{Right panel:}
    Same as in the left panel but for the charged chemical potential.}
\end{figure*}

Note that the time-averaged charge fraction measured during the whole
simulation is larger than $Y_Q=0.05$, as shown in the left panel of
Fig.~\ref{Yq-mub}. This panel is similar to the left panel
Fig.~\ref{T-mub}, in the sense that it also follows the evolution of the
densest and hottest points of the merger simulation, but it relates the
charge fraction and the baryon chemical potential. Overall, the charge
fraction achieved is larger for larger chemical potentials, going up to
$Y_Q\simeq0.12$, right before the temperature starts to increase. When
this happens, $Y_Q$ drops as a result of the appearance of the quarks
(medium-green diamonds). This occurs before the phase transition takes
place for the densest points. For the hottest points, $Y_Q$ increases at
the deconfinement phase transition (orange to brown circles), beyond
which they they correspond to the stellar center.

Since electric charge neutrality is always required for stellar
stability, the charge density of electrons has to balance the charge
density of baryons and quarks. As a consequence, the lepton fraction
$Y_e$, defined as the number of electrons over the number of baryons and
quarks, is the same as the charge fraction
\begin{eqnarray} Y_e = \frac{L}{B} = \frac{ n_e}{\sum_i Q_{B,i} \
n_i}=\frac{\sum_i Q_i \ n_i}{\sum_i Q_{B,i} \ n_i}=Y_Q . \label{Y_e}
\end{eqnarray} 
It was found in Ref.~\cite{Perego2019} using several hadronic equations
of state that the electron fraction does not go above $Y_e=0.12$ in
neutron star mergers, the same limiting value we obtained.

The right panel of Fig.~\ref{Yq-mub} again follows the evolution of the
densest and hottest points of the merger simulation, but now relates the
charged chemical potential and the baryon chemical potential. It is
interesting to note that, separately in each phase, the relation between
the two quantities is approximately linear (the light-green diamonds
present a slightly different slope because they represent a cold region
with no quarks). The very different slopes at different times (top and
bottom of panel) stem from the fact that the charged chemical potential
increases (in absolute value) with density much faster in the hadronic
phase than in the quark one. This behaviour has already been shown in
Ref.~\cite{Roark2018} for both charged and lepton chemical potentials
for the particular case of fixed temperature and in chemical
equilibrium. The jump from the bottom to the top line points to the
first-order deconfinement transition that takes place in the simulation.

\begin{figure}[t!] 
  \includegraphics[width=8.7cm, trim=0 0 0
    1.7cm,clip]{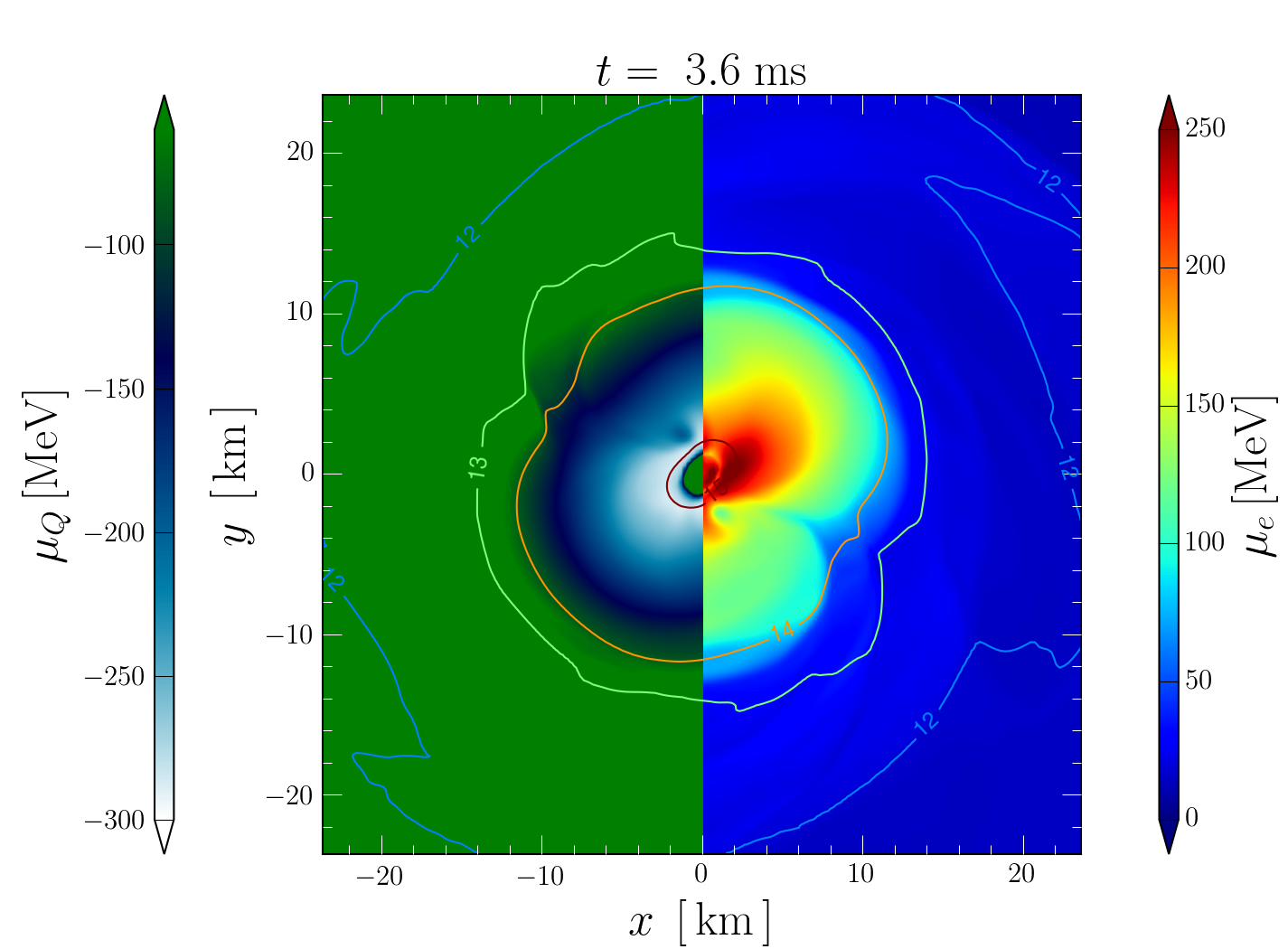}
  \caption{\label{muq_HMNS__046}Snapshot of the hypermassive neutron star immediately after
     deconfinement to quark matter has taken place in its
    center showing the charge chemical potential (left part) and electron
    chemical potential (right part). Contours refer to the
    rest-mass density.}
\end{figure}

Next, we concentrate again on a specific (same as in Fig.~\ref{T-mubv1}) time during the simulation after the deconfinement to quark
matter has taken place to discuss how the different chemical
potentials are spatially distributed within the HMNS. The contours in
Figs. \ref{muq_HMNS__046}--\ref{strange_HMNS__046} refer to values of the
rest-mass density boundaries between $10^{12}-10^{15}\,\rm g / cm^3$, the
latter being equivalent to $0.6$ fm$^{-3}$. The left part of Fig.~\ref{muq_HMNS__046} shows the charged
chemical potential. It can be seen how it increases (in absolute value)
with density towards the center of the HMNS until the phase transition
takes place, when it decreases rapidly
(in absolute value). The right part of Fig.~\ref{muq_HMNS__046}, on the
other hand, reports the electron chemical potential and shows that it
increases (on average) toward the center of the HMNS, being almost not sensitive to the
deconfinement phase transition. The difference between these two
quantities can be seen as a measure of the distance from chemical
equilibrium, when by construction $\mu_e=-\mu_Q$. 

At zero temperature, we can use the definition of the number density
$n_i=({\gamma_i}/{6\pi^2})k_{F_i}^3$ and of the Fermi momentum
$k_{F_i}=\sqrt{(\mu_i+\rm{vec} ^2-{M_i^*}^2)}$,
together with Eqs.~\eqref{mu_i} and \eqref{Y_e} to write a general
relation connecting electron and charged chemical potential
\begin{eqnarray} \mu_e =&&\Bigg[\left(\sum_i \frac{\gamma_i}{\gamma_e} Q_i
  [(Q_{B_i}\mu_B+Q_i\mu_Q+\mathrm{vec})^2-M{^*_i}^2]^{3/2}
\right)^{2/3}\nonumber\\&&\phantom{\Bigg[\biggl(}+m_e^2\Bigg]^{1/2} . \end{eqnarray} 
We used ``vec'' referring to general vector interactions, but they are specified in Eq.~(4) for our model. In the case in which both quark and hyperon degrees of freedom are
suppressed, the expression above reduces to
\begin{eqnarray} \mu_e
  &=&\Big[(\mu_B+\mu_Q+\mathrm{vec})^2-M{^*_p}^2+m_e^2\Big]^{1/2} .
\end{eqnarray} 

The left part of Fig.~\ref{st_HMNS__046} shows how the entropy density
per baryon number density $S_B=s/n_B$ is distributed in space inside the HMNS at the same
time of Fig.~\ref{muq_HMNS__046}. On average, it decreases towards the
center, reaching values close to $S_B=1$ right before the deconfinement
phase transition takes place, beyond which it increases to
$S_B\gtrsim2.5$. The right part of Fig.~\ref{st_HMNS__046} shows instead
the temperature distribution, highlighting that, at intermediate
densities, a hot ring appears around the center of the HMNS (see
Ref.~\cite{Hanauske2016} for an extended discussion of this
feature). More importantly, the center of the HMNS becomes very hot at
this point in time, i.e., with temperatures $T>60$ MeV. This is
consequence of the gravitational collapse we already discussed. Globally, the entropy $S_B$ increases as the simulation evolves in time, which is related to the appearance of shock waves that develop due to the large compressions experienced by matter as the HMNS settles. This is in a way similar to the dynamics produced in heavy-ion collisions
\cite{Bouras2009,Bouras2010} or in supernova explosions \cite{Burrows83}.

The amount and location of exotic particles (not nucleons) can be seen in
Fig.~\ref{strange_HMNS__046} for the same time as in the previous
figures. More specifically, the left part reports the baryon and quark
strangeness fraction
\begin{eqnarray}
Y_S=\frac{S}{B}=\frac{\sum_i Q_{S_i}}{\sum_i Q_{B,i}} ,
\end{eqnarray}
showing that it increases almost continuously towards the center of the
HMNS, reaching $Y_S\simeq40\%$. This fraction is composed mainly of
hyperons until the phase transition, when they are replaced by strange
quarks. The right part of Fig.~\ref{strange_HMNS__046} shows how the
strange quark fraction is present in low numbers and only in the hot
region before the deconfinement transition, but increases reaching
$Y_{\rm s-quark}\simeq40\%$ in the HMNS center.

\section{Comparison with other scenarios: supernovae and
  heavy-ion collisions}
  
\begin{figure}[t!] 
  \includegraphics[width=8.7cm, trim=0 0 0 1.7cm,clip]{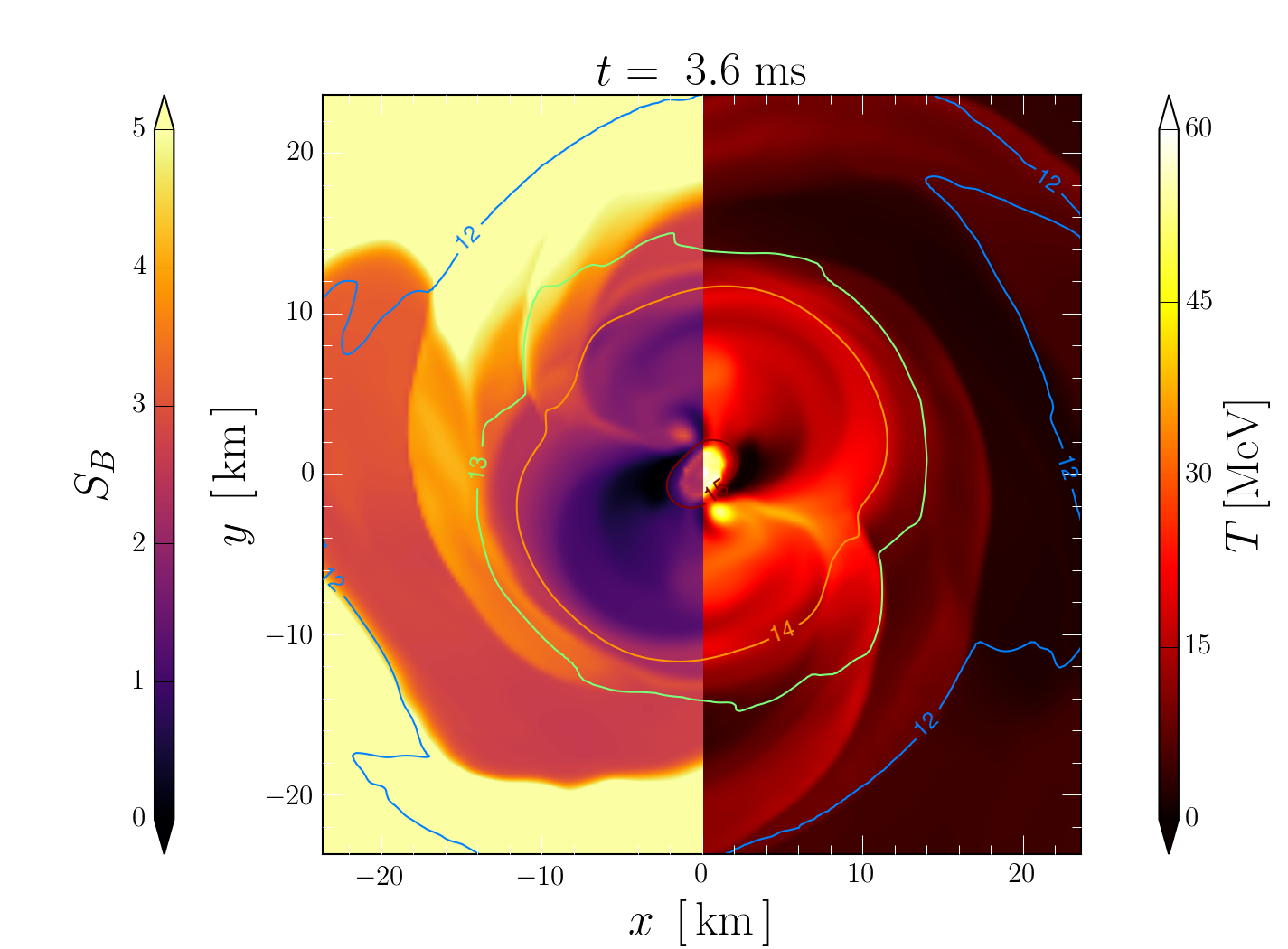}
  \caption{\label{st_HMNS__046}Same as Fig.~\ref{muq_HMNS__046} but
    showing entropy density per baryon density (left part) and
    temperature (right part).}
\end{figure}  

The physical conditions discussed so far and potentially encountered
after the merger of a binary system of neutron stars could be produced
also in two other and different scenarios, namely, supernovae explosions
and relativistic heavy-ion collisions.

Right after supernova explosions, the hot and dense medium of young
protoneutron stars causes the mean free path of neutrinos to drop
dramatically and becoming smaller than the radius of the star
\cite{Prakash97, Reddy1997, Shen:2003ih, Pastore:2014yua}. This is
usually modeled in nuclear physics assuming a large electron lepton
fraction (electron and electron neutrinos) with an electron fraction up
to $Y_e=0.4(=Y_Q)$, a value obtained from numerical simulations of
protoneutron-star evolution \cite{Fischer2010,Huedepol2010}. From
Fig.~\ref{Yq-mub}, it can be seen that the values reached in our merger
simulations are much smaller ($\sim 1/3$ of the typical protoneutron-star
value), and decrease further (to $\sim 1/6$ of the protoneutron-star
value in the center of the HMNS after the first-order phase transition
has taken place.

Similarly, the temperature evolution in protoneutron stars is usually
approximated by means of a fixed entropy density per baryon number
density (or entropy per baryon) $S_B$. When the entropy per baryon is
fixed in a EOS, it allows the temperature to increase almost linearly
with density (e.g., towards the stellar center), with typical values of
$S_B \simeq 1-2$ \cite{Prakash97,Pons1999}, again as deduced from
numerical simulations of protoneutron-star
evolution. Fig.~\ref{st_HMNS__046} shows that the values reached in our
merger simulations are comparable with those typically encountered in
protoneutron-star, being only a bit higher, $S_B=2.6$, in the region
where the first-order phase transition has taken place.  Note, however,
that in a supernova explosion in which quark deconfinement takes place,
the entropy per baryon can be even higher than in our case, reaching
$S_B=3$ \cite{Fischer2017}.

In our simulations, as it is typical of matter in astrophysical
scenarios, the net strangeness is nonzero, i.e., amount of strange
particles is larger than that of strange antiparticles. This is not the
case for matter generated in relativistic heavy-ion collisions, since in
this case there is no time for net strangeness to be produced, and a new
constraint needs to added to the formalism $Y_S=0$ \cite{Hempel:2009vp,
  Hempel:2013}. Note, however, that in a heavy-ion collision large
numbers of pions and kaons can and do escape from the system's surface
during the expansion phase, enriching strangeness. As a consequence, at
the final chemical freezeout, the strangeness fraction can reach rather
high values in the collision core made up of quarks and gluons
$Y_{S}\simeq 0.7$. This is to be contrasted with the maximum value
obtained in our merger simulation $Y_{S} \simeq 0.40$. Notwithstanding
this difference and and the fact that no electrons are involved in
heavy-ion collisions, we draw some rough comparisons. For instance,
typical Au-Au and Pb-Pb (and even U) collisions create environments with
charge fraction $Y_Q \simeq 0.4$, which is once more much larger than the
maximum value produced in our merger simulations, i.e., $Y_Q \simeq
0.12$.

\begin{figure}[t!] 
  \includegraphics[width=8.7cm, trim=0 0 0 1.7cm,clip]{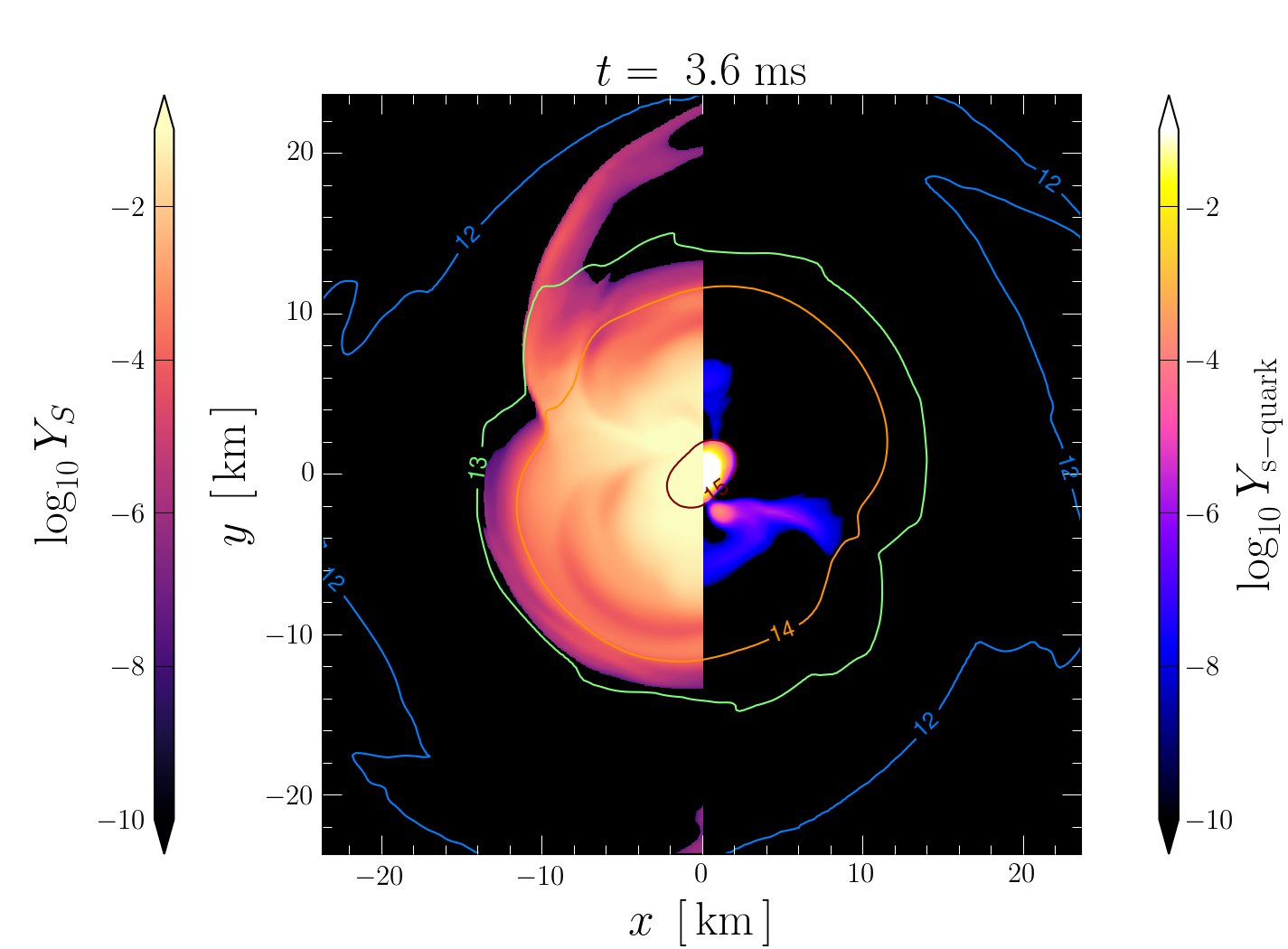}
  \caption{\label{strange_HMNS__046}Same as Fig.~\ref{muq_HMNS__046} but
  showing strangeness fraction  $Y_{\rm S}$ (left part) and strange-quark
  fraction $Y_{\rm s-quark}$ (right part).}
\end{figure}

Finally, low-energy collisions with energy per nucleon pair
$\sqrt{s_{NN}}<3$ GeV are expected to produce densities of several times
saturation density. These densities are even beyond the scope of the second phase of the Beam Energy Scan performed in the Relativistic Heavy Ion Collider at the Brookhaven National Laboratory (RHIC BES-II), but will be the focus of the Facility for Antiproton and Ion Research at the Gesellschaft fur Schwerionenforschung in Germany (GSI FAIR) and
the Nuclotron-Based Ion Collider Facility in Russia (NICA). In
particular, energies per nucleon pair of $\sqrt{s_{NN}}=2.2, 2.4, 2.6$
GeV are expected to generate initial-state temperatures of $T=60, 70, 80$
MeV, respectively, hence, corresponding to entropies per baryon of $S_B=3,
3.5, 4$ (see Tab.~II and discussion in Ref.~\cite{Motornenko:2019arp} for
details). Some of these temperatures and entropies are comparable or only
marginally above what we found in our merger simulations. 

\section{Discussion and conclusions}

We have recently presented the first fully general-relativistic
simulations showing that quark deconfinement can generate observable
signatures in the gravitational waveforms from merging neutron-star
binaries \cite{Most2018b}. In order to better understand the details and
the impacts of the deconfinement first-order phase transition, we have
discussed here a number of phase diagrams illustrating the properties of
the resulting hypermassive neutron star (HMNS) in terms of the evolution
of temperature, baryon chemical potential, charged chemical potential,
and charge fraction.

In particular, we have shown which parts of the phase diagram can be
probed in a representative neutron-star merger that generates a $\simeq
2.9\,M_\odot$ HMNS in which a deconfinement to quark matter takes place
and indicated the approximate amounts of light and strange quark
fractions that are created in this way. In turn, this has revealed that a
considerable amount of strange quarks, i.e., $Y_{\rm s-quark} \simeq
40\%$, can appear in the hot center of the HMNS after the deconfinement
phase transition has taken place. Of course, this does not mean that
strangeness is not present before the actual phase transition, since
light quarks can appear in small amounts in earlier stages of the
merger. In addition, hyperons, are generated in large amounts before the
transition, as a result of the increase in density and
temperature. Therefore, even before the deconfinement phase transition,
strangeness fraction can reach $Y_S \simeq 40\%$.

Our study has also revealed that the charged fraction achieved in the
simulation decreases when the deconfinement takes place, not exceeding
$Y_Q\simeq 0.12$. This value is much smaller than that encountered in
typical protoneutron star calculations $Y_Q\simeq 0.4$ or the the ones
present in relativistic heavy-ion collisions, where again $Y_Q\simeq
0.4$. Furthermore, the charged chemical potential decreases (in absolute
value) when the transition takes place and does not match the value of
the electron chemical potential. This difference highlights the fact that
the merged system is far from chemical equilibrium. Very informative are
also the spatial distributions of the various thermodynamical quantities
at a representative time after the deconfinement phase transition has
taken place. We have illustrated that, while the charged chemical
potential is reduced dramatically across the phase transition, the
chemical potential of the electrons is not affected significantly.

Finally, we note that the study of all these key thermodynamical
quantities is useful to validate whether the physical conditions produced
in neutron-star mergers are indeed similar to the those generated in
other physical scenarios, such as supernova explosions or relativistic
heavy-ion collisions \cite{Most2018}. More specifically, we have found
this analogy to hold reasonably well when comparing our temperatures and
entropies with the conditions encountered in the matter ejected in
supernova explosions or in heavy-ion collisions at low energies, such as
the ones to be produced in FAIR and NICA. On the other hand, both the
supernova and the heavy-ion collision scenarios are not able to reproduce
the extremely high baryon chemical potentials (i.e., $\mu_B>1.5$ GeV,
equivalent to more than 10 times saturation density) that can be achieved in
neutron-star mergers. In this way, neutron-star mergers can provide a
unique piece to the understanding of matter at extreme conditions of
density and temperature.

As a concluding remark, we note that soon after our results were
presented \cite{Most2018b}, Aloy et. al. \cite{Aloy2019} presented
interesting results from a systematic investigation of the convexity of
equations of state (EOSs)\footnote{We recall that fluids following convex
  EOSs are such that their fluid elements increase their specific volume
  and decrease their pressure when overtaken by a rarefaction wave (i.e.,
  rarefaction waves are ``expansive''); similarly, they are compressed
  when overtaken by a compression wave (i.e., compression waves are
  ``compressive'').  Conversely, fluids following non-convex equations of
  state are such that their fluid elements behave rather ``anomalously'',
  that is, they are compressed by rarefaction waves and rarefied by
  compression waves \cite{Rezzolla_book:2013}.}. In particular, they
showed that non-convex thermodynamics, which can appear if the adiabatic
index decreases sufficiently rapidly with increasing density, can affect
the equilibrium structure of stable compact stars, as well as the
dynamics of unstable neutron stars. In the latter case, a compression
shock can be formed at the inner border of the convex region and affect
the gravitational collapse to a black hole, leaving imprints on
gravitational waveforms, which would be of increased amplitude
\cite{Aloy2019}. This result was shown to hold when using our Chiral Mean
Field (CMF) EOS, but also other EOSs that did not include a deconfinement
phase transition. More importantly, it was found in Ref.~\cite{Aloy2019}
that those EOSs developing non-convex thermodynamics without a
deconfinement first-order phase transition had a non-consistent treatment
of matter constituents (non-relativistic instead of relativistic) or used
specific sets of parameter in their relativistic mean-field approach that
resulted in unphysical properties. A similar conclusion has been drawn by
Schneider et. al in Ref.~\cite{Schneider:2019vdm}, who found that a
pion-condensation transition does not mimic a quark deconfinement phase
transitions, as it is usually less extreme and does not generate a second
neutrino burst in supernova explosions. When taken together, these
arguments strengthen our conclusion that a deconfinement phase transition
can indeed leave distinguishable observable signals in different stages
of neutron-star evolution.

\section*{Acknowledgements} 
Support for this research comes in part from PHAROS (COST Action
CA16214), the LOEWE-Program in HIC for FAIR, the European Union's Horizon
2020 Research and Innovation Programme (Grant 671698; call FETHPC-1-2014,
project ExaHyPE), the ERC Synergy Grant ''BlackHoleCam: Imaging the Event
Horizon of Black Holes'' (Grant No. 610058), and the National Science
Foundation under grant PHY-1748621. HS also acknowledges the Judah
M.-Eisenberg-Laureatus Professorship at the Fachbereich Physik at Goethe
University. The simulations were performed on the SuperMUC cluster at the
LRZ in Garching, on the LOEWE cluster in CSC in Frankfurt, and on the
HazelHen cluster at the HLRS in Stuttgart.

%
\bibliographystyle{apsrev4-1} 
\bibliography{aeireferences}
%
%
%

\end{document}